\documentclass[a4paper,fleqn,usenatbib]{mnras}

\usepackage[T1]{fontenc}
\usepackage{ae,aecompl}
\usepackage{ulem}

\usepackage{graphicx}	% Including figure files
\usepackage{amsmath}	% Advanced maths commands
\usepackage{amssymb}	% Extra maths symbols
\usepackage{multicol}        % Multi-column entries in tables
\usepackage{bm}		% Bold maths symbols, including upright Greek
\usepackage{pdflscape}	% Landscape pages

\title[Discovery of Two New Pulsars in 47\,Tucanae (NGC 104)]{Discovery of Two New Pulsars in 47\,Tucanae (NGC 104)}

\author[Z. Pan, et al.]{
Z. Pan,$^{1}$\thanks{E-mail: panzc@bao.ac.cn}
G. Hobbs,$^{2}$
D. Li,$^{1,3}$
A. Ridolfi$^4$,
P. Wang,$^{1}$
and P. Freire$^4$
\\
$^{1}$National Astronomical Observatories, Chinese Academy of Sciences, A20 Datun Road, Chaoyang District, Beijing 100012, China.\\
$^{2}$CSIRO Astronomy and Space Science, CSIRO, PO Box 76, Epping, NSW 1710, Australia\\
$^{3}$Key Laboratory of Radio Astronomy, Chinese Academy of Sciences, Nanjing 210008, China \\
$^{4}$Max-Planck-Institut f{\"u}r Radioastronomie MPIfR, Auf dem H{\"u}gel 69, D-53121 Bonn, Germany
}

\date{Accepted XXX. Received YYY; in original form ZZZ}

\pubyear{2015}

\begin{document}
\label{firstpage}
\pagerange{\pageref{firstpage}--\pageref{lastpage}}
\maketitle

\begin{abstract}
We report the discovery of two new millisecond pulsars (PSRs J0024$-$7204aa and J0024$-$7204ab) in the globular cluster 47\,Tucanae (NGC 104).
Our results bring the total number of pulsars in 47\,Tucanae to 25.
These pulsars were discovered by reprocessing archival observations from the Parkes radio telescope.
We reprocessed the data using a standard search procedure based on the PRESTO software package
as well as using a new method in which we incoherently added the power spectra corresponding to $\sim$1100\,hr of observations.
The newly discovered PSR~J0024$-$7204aa, has a pulse frequency of $\rm \sim$541\,Hz (corresponding to a $\rm \sim$1.84 ms period),
which is higher than any other pulsars currently known in the cluster and ranks 12$^{\rm{th}}$ amongst all the currently known pulsars.
The dispersion measure of this pulsar, 24.941(7)\,cm$^{-3}$ pc, is the highest in the cluster.
The second discovered pulsar, PSR~J0024$-$7204ab, is an isolated pulsar with a pulse frequency of $\rm \sim$270\,Hz (corresponding to a period of $\rm \sim$3.70 ms).
\end{abstract}


%%%%%%%%%%%%%%%%%%%%%%%%%%%%%%%%%%%%%%%%%%%%%%%%%%

%%%%%%%%%%%%%%%%% BODY OF PAPER %%%%%%%%%%%%%%%%%%

\section{Introduction}

Since the discovery of the first globular cluster (GC) pulsar,
PSR~B1821$-$24A in M28 (Lyne et al. 1987),
GC environments have proven to be conducive to pulsar searches.
Before January 2016, 144 radio pulsars in 28 GCs\footnote{http://www.naic.edu/$\rm \sim$pfreire/GCpsr.html} had been discovered.
Many of the pulsars are exotic.
With a spin frequency of $\rm \sim$716\,Hz,
PSR~J1748$-$2446ad is the fastest-spinning pulsar known (Hessels et al.\ 2006).
That pulsar was discovered in the globular cluster Terzan 5,
which has 35 known pulsars (Lyne et al.\ 1990; Lyne et al.\ 2000; Ransom et al.\ 2005; Hessels et al.\ 2006; GC pulsars catalog$^{1}$), the largest for any GCs.
Among all the GC pulsars, 80 are known to be in binary systems and 132 are millisecond pulsars with spin periods of 30 ms or shorter.
Both fractions are much higher than those of the Galactic population (see psrcat, Manchester et al.\ 2005).

%The GC 47\,Tucanae was known to contain 23 radio pulsars.
%This cluster has been observed using telescopes in the Southern Hemisphere such as the Australia Telescope Compact Array (e.g., McConnell et al. 2004) and more regularly using the 64-m Parkes radio telescope.
With 23 previously known radio pulsars, 47\,Tucanae (NGC 104) is the second most populated cluster.
All the pulsars were discovered using the 64-m Parkes radio telescope (Manchester et al.\ 1990; Manchester et al.\ 1991; Robinson et al.\ 1995; Camilo et al.\ 2000; Freire et al.\ 2016, in preparation) and are millisecond pulsars.
Freire et al. (2001a) used observations of the pulsars in 47\,Tucanae to make the first detection of ionized intracluster gas.
Timing models for 15 of the pulsars in the cluster were presented by Freire et al. (2001b) and updated by Freire et al. (2003).
%More recently the pulsars in the cluster were shown to have association with near-ultraviolet companions (Rivera-Sandoval et al. 2015 and Cadelano et al. 2015).

One of the primary goals of modern-day pulsar astronomy is to discover a millisecond pulsar in orbit around a stellar mass black hole.
Evolutionary models suggest that such a system is only likely to form in environments with high stellar interaction rates.
Therefore GCs provide some of the most likely environments in which to find such a system.
Finding such exotic systems is a primary goal of the Five-hundred-meter Aperture Spherical radio Telescope (FAST) (Nan et al. 2011)
currently under construction in China. % and is expected to start searching for pulsars around September 2016.
The full system testing will start in the middle of June 2016 followed by an expected first light on September 26$^{\rm{th}}$.
Zhang et al. (submitted to RAA) described how early FAST observations of GCs may find pulsars by carrying out repeated short drift-scan observations.
Traditional pulsar searches are based on the analysis of a single, long time sequence of data.
The requirement to carry out short observations during early FAST operation implies that the data from multiple observations (taken on different days) will need to be combined.
%To obtain experience in combining such data sets we made use of a large number of archival observations of 47\,Tucanae obtained using the Parkes radio telescope.
%As described below our processing allowed us to study pulsar observation data combination and related pulsar search that will occur with the future FAST observations,
%but also allowed us to discover two new pulsars in the cluster.
We thus developed a segmented search method operating on power spectra.
The testing and application of such a method on the large number of 47\,Tucanae observation data allowed us to discover two new pulsars.
Such experience is also valuable when planning our future FAST pulsar surveys.

In Section 2 of this paper, we describe the details of the observations and data reduction.
In Section 3, we describe the discovery of two new pulsars.
The conclusions are in Section 4.

\section{Observations and Processing}

The aim of the Commonwealth Scientific and Industrial Research Organization (CSIRO) pulsar data archive\footnote{http://data.csiro.au} (Hobbs et al.  2011)
is to store pulsar data sets obtained using the Parkes radio telescope.
The archive now contains more than 300\,TB of data and we were collating as much archival data as possible.
Freire et al. (private communication) provided copies of observations of 47\,Tucanae obtained between the years 1999 and 2011.
The observations were described in Freire et al. (2001a) and Freire et al. (2001b).
An analogue filterbank backend was used to record the data with 512 frequency channels and 0.5\,MHz channel width, corresponding to a total bandwidth of 256\,MHz centered at 1390\,MHz.
The output of each filter was integrated for 80 $\mu$s and the resulting power was then 1$-$bit digitized and stored onto magnetic tapes.
The majority of the observations were obtained with the central beam of the 13-beam multibeam receiver (Staveley-Smith et al.\ 1996) in the 20\,cm band.
The integration times varied from a few minutes (for test observations) to $\sim$8\,hr.
The dataset consisted mostly of high time resolution data (aforementioned 80 $\mu$s sampling time) and a small fraction of lower time resolution data (125 $\mu$s sampling time).

The raw data files were originally extracted from the tape files using the \textsc{sc\_td} software package and
then converted to the SIGPROC\footnote{https://github.com/SixByNine/sigproc/} filterbank format using the \textsc{filterbank} procedure.
The data were processed using two methods described below.

The first method was a traditional pulsar acceleration search using the PRESTO\footnote{http://www.cv.nrao.edu/$\rm \sim$sransom/presto/} suite of software (Ransom et al.\ 2002).
This method was applied to the observations individually.
We initially de-dispersed the data with a dispersion measure (DM) corresponding to about the median of the values of the known pulsars in the globular cluster, namely 24.5\,cm$^{-3}$ pc, and then summed across the frequency channels.
We then used the \textsc{realfft} and \textsc{accelsearch} routines to perform the pulsar search.
%For the PRESTO acceleration search, we searched with a zmax value (the -zmax value, which is an option for the routine \textsc{accelsearch}) of 50 for possible binary pulsar signals and summed up to 16 harmonics.
%For the PRESTO acceleration search,
%we summed up to 16 harmonics and chose 50 as the max value of $z$ (the -zmax option) which was the number of Fourier bins that the pulsar frequency drifted over the course of an integration (Ransom et al. 2001).
%The average acceleration of the pulsar in the integration can be calculated as $a = z \rm{c}/(T^2 f_0)$ (Ransom et al. 2001),
%in which c is the speed of light, $T$ is the total integration time, $f_0$ is the pulsar spin frequency.
%For our observation data, if the total integration time was one hour, using 50 as the zmax value corresponded to an acceleration of 5.8 m s$^{-2}$ for a 200 Hz signalor 2.3 m s$^{-2}$ for a 500 Hz signal.
For the PRESTO acceleration search,
we summed up to 16 harmonics and chose 50 as the maximum value of $z$ (through the -zmax option in \textsc{accelsearch}).
This latter corresponds to the maximum number of adjacent Fourier bins that \textsc{accelsearch} considers as the result of a possible drift of the intrinsic pulsar spin frequency, due to the binary acceleration\footnote{The average acceleration of the pulsar in the integration can be calculated as $a = z \rm{c}/(T^2 f_0)$ (Ransom et al. 2001), in which c is the speed of light, $T$ is the total integration time, $f_0$ is the pulsar spin frequency.
For our observation data, if the total integration time was one hour, using 50 as the zmax value corresponded to an acceleration of 5.8 m s$^{-2}$ for a 200 Hz signal or 2.3 m s$^{-2}$ for a 500 Hz signal.} (Ransom et al. 2001).

For the second method we segmented each observation into small files using the \textsc{extract} tool in the SIGPROC software package.
Each file contained $2^{23}$ samples corresponding to an observation lasting $\rm \sim$11 minutes.
In this process we ignored observations shorter than 11 minutes and also discarded any data that remained after dividing the observation into these residual segments.
From all the observation data, we ended up with 5956 segmented files corresponding to $\sim$1110\,hr of observation.
Each of these 5956 segmented files were then de-dispersed (again using a dispersion measure of 24.5\,cm$^{-3}$ pc) and summed across the frequency channels.
During the de-dispersion, some samples were lost due to the DM delay.
For the 5956 segmented files, in total we lost $\rm \sim$243 second observation data, which is much shorter than the total length of $\rm \sim$1110\,hr.
We then stored the Fast Fourier Transform (FFT) of each of these time series on disk.
We repeated this procedure, but did not de-disperse the data files, i.e., this final processing corresponded to a dispersion measure of 0\,cm$^{-3}$ pc.

\begin{figure}
%\centering
 \hspace{-3em}
 \vspace{-1em}
 \includegraphics[width=70mm,angle=90]{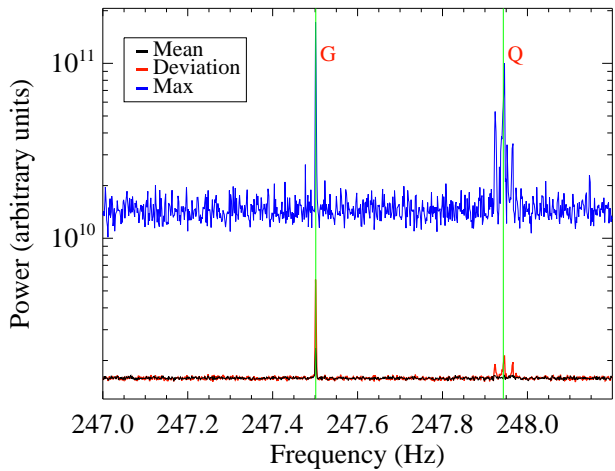}
 \caption{%Frequency domain data around the frequency of previous known pulsar J0024-7204G and J0024-7204Q (a binary pulsar). The green vertical lines show the their rotation frequencies.}
 Frequency domain data around the spin frequencies (green vertical lines) of the previously known pulsars J0024$-$7204G (isolated) and J0024$-$7204Q (binary).}
 \label{g_q}
\end{figure}

Note that we did not expect to detect a given pulsar in each data segment.
The pulsars in 47\,Tucanae are known to scintillate and some of the binary systems can eclipse the pulsar signal (such as, e.g., eclipsing PSRs J0024$-$7204J and J0024$-$7204O, described in Freire et al. 2003).
The first method produced a list of pulsar candidates for each observation that were then inspected by eye.
Analysing the FFTs from the second method was more complex and carried out as follows.
%For each spectral frequency channel we had the real and imaginary parts of the Fourier transform for each observation segment.
%These were summed in quadrature to form the power in that frequency channel for each observation segment.
%We then calculated, for each spectral channel, the mean, maximum and standard deviation of the power values over all the observation segments.
For each spectral frequency channel of every observation segment, we had the real and imaginary parts from the Fourier transform.
The real and imaginary parts were summed in quadrature to form the power values.
From 5956 observation segments, we obtained 5956 power values for every frequency channel.
For all the power values of the same frequency channel, the mean value, the maximum value, and the standard deviation were calculated.
An example is shown in Figure~\ref{g_q}, that shows the power spectra around the known pulse frequencies for PSRs J0024$-$7204G and Q (indicated by the vertical green lines).
The black line is the mean of the power in all the observation segments.
PSR~J0024$-$7204G is clearly visible in the mean values, but Q is not.
This is because the flux density of Q varies significantly (likely to be caused by scintillation) and so the mean flux density is low.
In constrast, both pulsars are easily detectable in either the maximum power value or the standard deviation (the blue and red lines respectively).
Note that Q is in a binary system and so the signal is spread over multiple spectral channels.

%Unfortunately,
%radio-frequency-interference (RFI) leads to spurious signals in the mean, maximum and standard deviation power spectra and
%therefore we compare (by eye) the same frequency range obtained using the data that has not been de-dispersed (the zero-DM spectra).
%We record (as possible pulsar candidates) any signal that is detectable in the de-dispersed spectra,
%but not in the zero-DM spectra\footnote{Note that if we filter the spectra in order to only display frequency channels corresponding to pulse periods longer than 0.5\,ms then we must inspect 1342173 channels.  If we show 4096 %channels simultaneously then we require that 328 figures are inspected for pulsar candidates. This can easily be carried out by eye by a single person.}.

Unfortunately,
the statistical behaviours of radio-frequency-interferences (RFIs) can and do mimic those of pulsars.
To weed out spurious signals caused by RFIs,
we inspected by eye the zero-DM spectrum (having not been de-dispersed) corresponding to each candidate.
Note that if we filter the spectra in order to only display frequency channels corresponding to pulse periods longer than 0.5\,ms then we must inspect 1342173 channels.
If we show 4096 channels simultaneously then we require that 328 figures are inspected for pulsar candidates. This can easily be carried out by eye by a single person.
We recorded, as possible candidates, only those signals without counterparts in the zero-DM spectrum.

\section{Results}

%Our methods re-detected all the known pulsars in 47\,Tucanae except for P, R, V, and W, which were missed by both procedures.
%These undetected pulsars are extremely faint and more details of those pulsars will be presented by Ridolfi et al. (in preparation)
%who used significantly more observations than those available to us.
Our segmented search re-detected all the known pulsars in 47\,Tucanae except for P, R, V and W, which were missed by both procedures.
These undetected pulsars are not only extremely faint,
but also very accelerated. This latter condition causes their signals to be significantly spread in the Fourier domain.
More details of those pulsars will be presented by Ridolfi et al. (in preparation) who used significantly more observations than those available to us.
Other four pulsars (S, T, U and Y) were only missed by our PRESTO acceleration search.
Missing these pulsars was due to our candidate selection settings.
%These four pulsars only occurred 5, 10, 14, and 12 times in all the PRESTO acceleration search result for the 820 original data files and were ignored as occasional RFIs.
In fact, the spin frequencies of pulsars S, T, U and Y were detected only 5, 10, 14 and 12 times, respectively, in all the PRESTO acceleration search result for the 820 original data files and were ignored as occasional RFIs.

%In order to search for new pulsars we first analysed the segmented search method.
Inspecting signals from the segmented search result led to 28 candidates from the mean power spectra, 392 from the maximum values, and 131 from the standard deviations.
Confirming these candidates is challenging.
Traditionally a further observation is carried out to try to re-detect the candidate.
However, the segmented search method made use of $\rm \sim$1100\,hr of observations - this cannot easily be reproduced.
To confirm or deny our candidates,
we therefore folded each individual data set using \textsc{prepfold} around the candidate period,
allowing a search in pulse period to maximise the signal-to-noise ratio of the resulting profile.
We then searched for multiple observations that exhibited profiles with similar shape and,
where possible, attempted to obtain a phase-connected timing solution. This process removed all but one of our segmented-search pulsar candidates.

\begin{figure}
\centering
\includegraphics[width=80mm,angle=0]{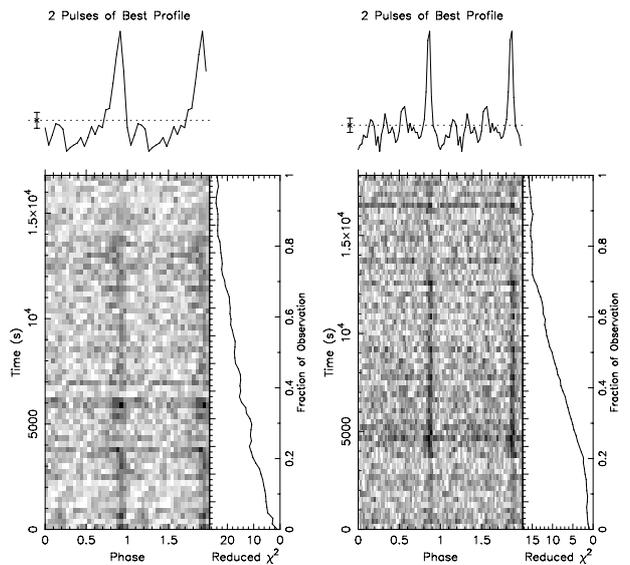}
\caption{Averaged pulse profile (top panels) and intensity versus phase and time (bottom panels) for PSR~J0024$-$7204aa (left) and PSR~J0024$-$7204ab (right),
as they appeared in their discovery diagnostic plots of PRESTO.}
\label{fg:aa_ab}
\end{figure}

We then inspected the PRESTO acceleration search method results, and found 1134 candidates in total.
These candidates were inspected by eye.
Among the 1134 candidates,
908 candidates were identified as RFI signals from their narrow band and/or zero-DM feature(s) and 46 candidates were signals from known pulsars.
The remainders were inspected by eye in more detail, and only two of these candidates were considered of sufficient quality for further investigation.
One of these corresponded to the same candidate as that found by the segmented search method.
Hereafter we label this pulsar as PSR~J0024$-$7204aa.
We label the other detection as PSR~J0024$-$7204ab.
We carefully checked to see whether the pulse frequency for the new pulsars were harmonically related to the known pulsars in the cluster,
but were unable to find any such match.
Figure~\ref{fg:aa_ab} shows the averaged profiles and phase-time plots of the discovery figures in which we found the two new pulsars.

\subsection{PSR~J0024$-$7204aa}

To date, we have been unable to obtain a timing solution for this pulsar, but have clearly detected it in 12 of the original data files.
This pulsar was detected by the PRESTO pipeline without exhibiting any evidence that it is accelerating.
Therefore, it is unlikely to be in a fast binary system.
It has a pulse period of $\rm \sim$1.84 ms,
which is shorter than the rotational periods of any other known pulsars in the cluster.
Its dispersion measure, determined by measuring pulse arrival times in different sub-bands of the brightest three available detections, is 24.971(7) cm$^{-3}$ pc,
which is currently the highest in the cluster.

\subsection{PSR~J0024$-$7204ab}

\begin{figure}
\centering
\vspace{-2em}
\includegraphics[width=65mm,angle=270]{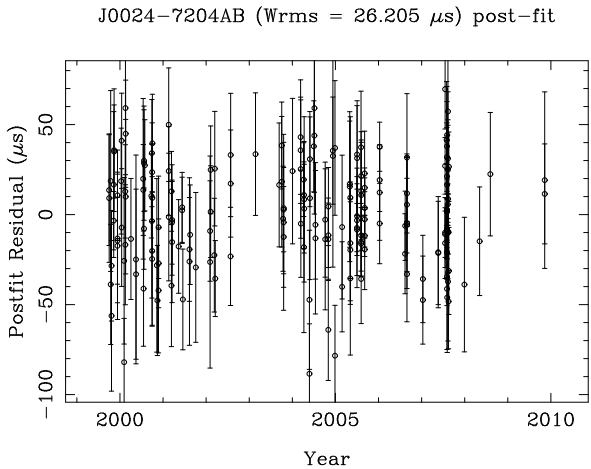}
\caption{Timing residuals for PSR~J0024$-$7204ab.}\label{fg:residuals}
\end{figure}

\begin{figure}
\centering
\includegraphics[width=80mm,angle=0]{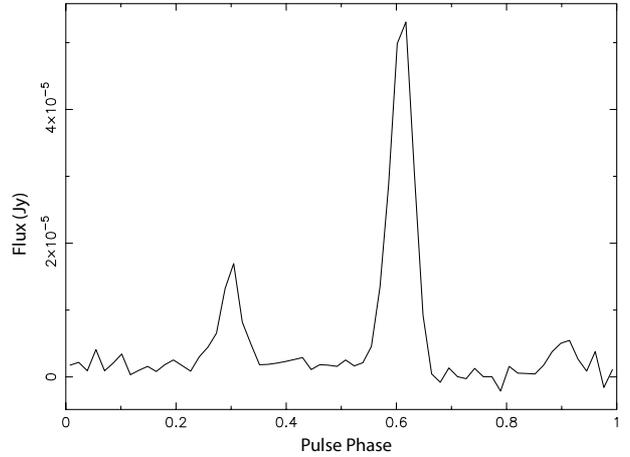}
\caption{Averaged pulse profile of PSR~J0024$-$7204ab obtained by coherently adding all the 87 detections of the pulsar found in our high time resolution (sampling time 80 $\mu$s) dataset.
The signal-to-noise ratio of this averaged pulse profile is 49.145. }
\label{fg:averaged_ab}
\end{figure}

%We have successfully obtained a timing solution using all the available data for PSR~J0024$-$7204ab.
Similar to PSR~J0024$-$7204aa, the dispersion measure was obtained by extracting TOAs from different sub-bands from the brightest detections (six, in this case).
We have then successfully obtained a timing solution using all the available data for PSR~J0024$-$7204ab.
The TEMPO2 (see e.g., Hobbs, Edwards \& Manchester 2006) solution is tabulated in Table~\ref{tb:psr_ab}.
The phase-connected timing residuals using this timing model are shown in Figure~\ref{fg:residuals}.
Adding all the 87 detections, we obtained the pulse profile of this pulsar (see Figure~\ref{fg:averaged_ab}).
There is a clear second peak at phase 0.3 and a possible third peak at phase 0.9.
The phase connected pulse arrival times allowed us to determine the position of the pulsar.
This is shown, in relation to the other pulsars in the cluster, in Figure 5.
The pulsar is clearly associated and situated in the central region of 47\,Tucanae.
Its proper motion is consistent with the global proper motion of the cluster as reported by Freire et al. (2003).

\begin{table}
\hspace{-2em}
\caption{Parameters for PSR~J0024$-$7204ab.}\label{tb:psr_ab}
\begin{footnotesize}\begin{tabular}{ll}\hline\hline
\multicolumn{2}{c}{Fit and Data-set} \\
\hline
Pulsar name\dotfill & J0024$-$7204ab \\
MJD range\dotfill & 51451.6---55143.5 \\
Data span (yr)\dotfill & 10.11 \\
Number of TOAs\dotfill & 192 \\
Rms timing residual ($\mu$s)\dotfill & 26.2 \\
Weighted fit\dotfill &  Y \\
Reduced $\chi^2$ value \dotfill & 1.1 \\
\hline
\multicolumn{2}{c}{Measured Quantities} \\
\hline
Right ascension, RA (hh:mm:ss)\dotfill &  00:24:08.1657(4) \\
Declination, DEC (dd:mm:ss)\dotfill & $-$72:04:47.616(2) \\
Pulse frequency (Hz)\dotfill & 269.931793752167(14) \\
%Pulse period (ms)\dotfill & 3.70463955393906(4)  \\
Pulse frequency derivative\dotfill & $-$7.1728(20)$\times 10^{-16}$ \\
Proper motion in RA (mas\,yr$^{-1}$)\dotfill & 4.3(6) \\
Proper motion in DEC (mas\,yr$^{-1}$)\dotfill & $-$2.6(6) \\
\hline
\multicolumn{2}{c}{Set Quantities} \\
\hline
Reference epoch (MJD)\dotfill & 53600 \\
Dispersion measure, DM (cm$^{-3}$ pc)\dotfill & 24.37(2) \\
\hline
\multicolumn{2}{c}{Assumptions} \\
\hline
Clock correction procedure\dotfill & TT(TAI) \\
Solar System ephemeris model\dotfill & DE421 \\
%Binary model\dotfill & NONE \\
Model version number\dotfill & 5.00 \\
\hline
\end{tabular}\end{footnotesize}
\end{table}

\begin{figure}
\centering
\includegraphics[width=80mm,angle=0]{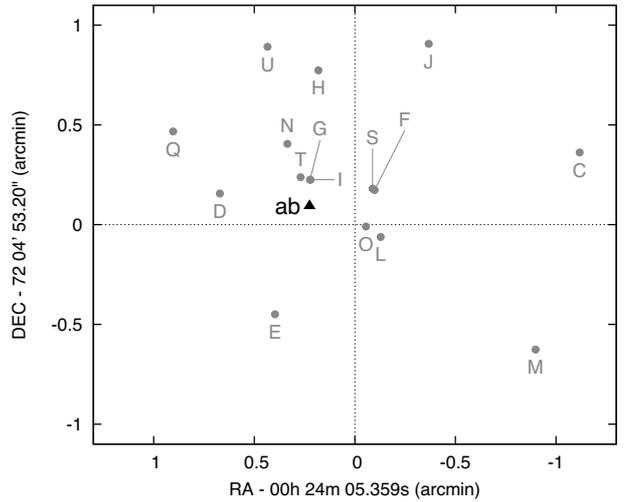}
\caption{The positions of PSR~J0024$-$7204ab and the other pulsars of 47\,Tucanae with known timing solutions.}
\label{fg:position_ab}
\end{figure}

\section{Conclusions}

  We have re-processed archival observations of 47\,Tucanae obtained with the Parkes radio telescope using two search algorithms
  and have re-detected the majority of the known pulsars
  and discovered two new millisecond pulsars.
  The \mbox{two} methods have different characteristics and can be complementary to each other.
  The traditional, PRESTO based method detected both of the new pulsars,
  but also led to a large number of false candidates. % that needed to be filtered out of the results.
  It also requires setting up processing pipelines, selecting the parameters for the search and was relatively slow. % on the computer systems that we had available.
  The new, segmented search, only detected one of the new pulsars (aa).
  The missing pulsar (ab) is weak and the segmented method subdivided the data into such short intervals that the signal-to-noise ratio of the pulsed signal was \mbox{$<$ 1} in each segment.
  However, the segmented search method easily detected relatively bright pulsars that are not in fast binaries without a large number of false candidates.
  As both methods are relatively straight-forward, we recommend using them both in future analyses of such data sets.

  The two new discoveries add to the growing population of pulsars in globular clusters.
  These two new pulsars are weak and only occasionally observable in the Parkes data sets.
  However, in the near future, much more sensitive telescopes, such as MeerKAT and the Square Kilometer Array (SKA), 
  will be operating in the Southern Hemisphere and will be able to make regular observations of these new pulsars.

\section*{Acknowledgements}

This work is supported by National Basic Research Program of China (973 program) No. 2012CB821800,
National Natural Science foundation of China No. 11373038,
and the Strategic Priority Research Program "The Emergence of Cosmological Structures" of the Chinese Academy of Sciences, Grant No. XDB09000000.
G.H. acknowledges the support from the professorship award under the Chinese Academy of Sciences (CAS) President's International Fellowship Initiative (PIFI) in 2015.   The Parkes radio telescope is part of the Australia Telescope,
which is funded by the Commonwealth of Australia for operation as
a National Facility managed by the CSIRO.
This paper includes archived data obtained through the CSIRO Data Access Portal.
We acknowledge that the data were obtained by numerous observers over many years.
We thank them for their support in sharing the data.
P.F. and A.R. gratefully acknowledge financial support by the European Research Council for the ERC Starting grant BEACON under contract No. 279702.
A.R. is a member of the International Max Planck research school for Astronomy and Astrophysics
at the Universities of Bonn and Cologne and acknowledges partial support through the Bonn-Cologne Graduate School of Physics and Astronomy.

%%%%%%%%%%%%%%%%%%%%%%%%%%%%%%%%%%%%%%%%%%%%%%%%%%



%\bibliographystyle{mnras}


%%%%%%%%%%%%%%%%%%%%%%%%%%%%%%%%%%%%%%%%%%%%%%%%%%

%%%%%%%%%%%%%%%%% APPENDICES %%%%%%%%%%%%%%%%%%%%%

%\appendix


%%%%%%%%%%%%%%%%%%%%%%%%%%%%%%%%%%%%%%%%%%%%%%%%%%

% Don't change these lines
\bsp	% typesetting comment
\label{lastpage}
\end{document}